\documentclass[aps, preprint, prl, superscriptaddress, showpacs,nobibnotes]{revtex4-1}
\usepackage{graphicx}
\usepackage{bm,amsmath}
\usepackage{dcolumn}
\usepackage[version=4]{mhchem}
\usepackage{natbib,xcolor}

\begin{document}
\title{A flat band-induced correlated kagome metal}

\author{Linda Ye}
\thanks{Present Address: Department of Applied Physics, Stanford University, Stanford, CA 94305, USA}
\affiliation{Department of Physics, Massachusetts Institute of Technology, Cambridge, MA 02139, USA}
\author{Shiang Fang}
\affiliation{Department of Physics and Astronomy, Rutgers University, Piscataway, NJ 08854, USA}
\author{Min Gu Kang}
\affiliation{Department of Physics, Massachusetts Institute of Technology, Cambridge, MA 02139, USA}
\author{Josef Kaufmann}
\affiliation{Institute for Solid State Physics, TU Wien, 1040 Vienna, Austria}
\affiliation{Institute for Theoretical Solid State Physics, Leibniz IFW Dresden, 01069 Dresden, Germany}
\author{Yonghun Lee}
\affiliation{Department of Physics, Massachusetts Institute of Technology, Cambridge, MA 02139, USA}
\author{Jonathan Denlinger} 
\affiliation{Advanced Light Source, Lawrence Berkeley National Laboratory, Berkeley, CA 94720, USA}
\author{Chris Jozwiak} 
\affiliation{Advanced Light Source, Lawrence Berkeley National Laboratory, Berkeley, CA 94720, USA}
\author{Aaron Bostwick} 
\affiliation{Advanced Light Source, Lawrence Berkeley National Laboratory, Berkeley, CA 94720, USA}
\author{Eli Rotenberg}
\affiliation{Advanced Light Source, Lawrence Berkeley National Laboratory, Berkeley, CA 94720, USA}
\author{Efthimios Kaxiras}
\affiliation{Department of Physics, Harvard University, Cambridge, MA 02138, USA}
\affiliation{John A. Paulson School of Engineering and Applied Sciences, Harvard University, Cambridge, MA 02138, USA}
\author{David C. Bell}
\affiliation{John A. Paulson School of Engineering and Applied Sciences, Harvard University, Cambridge, MA 02138, USA}
\affiliation{Center for Nanoscale systems, Harvard University, Cambridge, MA 02138, USA}
\author{Oleg Janson}
\affiliation{Institute for Theoretical Solid State Physics, Leibniz IFW Dresden, 01069 Dresden, Germany}
\author{Riccardo Comin}
\affiliation{Department of Physics, Massachusetts Institute of Technology, Cambridge, MA 02139, USA}
\author{Joseph G. Checkelsky}
\thanks{checkelsky@mit.edu}
\affiliation{Department of Physics, Massachusetts Institute of Technology, Cambridge, MA 02139, USA}
\date{\today}
\maketitle

\textbf{
The notion of an electronic flat band refers to a collectively degenerate set of quantum mechanical eigenstates in periodic solids \cite{Lieb,Mielke_Tasaki}. The vanishing kinetic energy of flat bands relative to the electron-electron interaction is expected to result in a variety  of many-body quantum phases of matter \cite{FB_SC,Sun_FB,Mudary_FB,Wen_FB}. Despite intense theoretical interest, systematic design and experimental realization of such flat band-driven correlated states in natural crystals have remained a challenge. Here we report the realization of a partially filled flat band in a new single crystalline kagome metal \ce{Ni3In}. This flat band is found to arise from the Ni $3d$-orbital wave functions localized at triangular motifs within the kagome lattice plane, where an underlying destructive interference among hopping paths flattens the dispersion. We observe unusual metallic and thermodynamic responses suggestive of the presence of local fluctuating magnetic moments originating from the flat band states \cite{FB_Local_Mag,FB_Local_Mag2}, which together with non-Fermi liquid behavior indicate proximity to quantum criticality \cite{QC_review}. These results demonstrate a lattice and orbital engineering approach to designing flat band-based many-body phenomena that may be applied to integrate correlation with topology \cite{Topo_Corr_review, Topo_Corr_review2} and as a novel means to construct quantum criticality \cite{NFL_review}.
}

Hopping of electrons across a periodic lattice introduces a band dispersion and quantum mechanical eigenstates characterized by itinerant, spatially delocalized Bloch waves (schematically depicted in Fig. 1(a)).  Under certain circumstances a given lattice can also support bands with little or no dispersion (see Fig. 1(b)) -- termed ``flat bands'' -- driven by competing terms within the hopping dynamics.  A key implication given a flat band dispersion is that the kinetic energy of electrons -- proportional to the band width -- is essentially quenched such that electronic states may be viewed as spatially localized (see inset of Fig. 1(b)).  If energetically brought to the Fermi level $E_F$, interaction effects in this highly degenerate electronic manifold can then give rise to many-body electronic states.  Decades of theoretical work on lattice-derived flat bands has suggested a number of intriguing possible ground states, including magnetic \cite{Lieb,Mielke_Tasaki}, superconducting \cite{FB_SC}, and fractionalized states \cite{Sun_FB,Mudary_FB,Wen_FB}, the latter predicted to host long-range entanglement akin to that in fractional quantum Hall \cite{FQHE} and quantum spin liquid states \cite{QSL_review}.

In terms of realization, flat bands were first discussed in lattice models with uniform hopping that satisfy certain connectivity conditions \cite{Balents_FB} leading to destructive quantum mechanical interference of electronic wave functions, exemplified by the dice \cite{Sutherland}, Lieb \cite{Lieb} and kagome \cite{kagome_Hoffman} networks. It has been recently demonstrated that the hopping pathways in these models can be constructed using synthetic quantum simulators \cite{ArtificalFB} as well as crystalline lattices \cite{Co3Sn2S2,CoSn}.  However, the study of interaction effects in the latter faces significant challenges, the principal of which being the stabilization of materials which favor the energetic position of the flat band near $E_{F}$.  Instead, apart from the rich correlation effects observed in Landau level flat bands \cite{FQHE}, it is primarily in recent reports of van der Waals moir\'{e} superlattices \cite{MacDonald_FB,MacDonald_TMD, twBLG, triG, twBLGferro,TMD_Moire} that such states have been observed. Whether a natural crystal can support flat band-driven correlated phenomena has remained an open question. In this letter we address this through the first realization and study of single crystals of a binary intermetallic compound \ce{Ni3In}, which leverages the $3d$-orbital degree of freedom to realize a Ni-based electronic band flat in the kagome plane near $E_F$.  We observe hallmarks of intense electronic correlation in the form of non-Fermi liquid behavior, which we propose to originate from unusual magnetic fluctuations associated with the flat band manifold.

Transition metal intermetallic compounds are a diverse and technologically important class of materials capable of supporting a wide variety of crystalline structures \cite{Alloy_1972,Intermetallics_2019}.  Combined with the $d$-electron orbital degrees of freedom, they afford a high degree of tunability for electronic structure which we employ for the realization of flat band dispersions.  A striking candidate as such is \ce{Ni3In} which crystallizes in space group $P6_{3}/mmc$. Its basic structural unit is a breathing Ni kagome network which circumscribes In in the hexagonal void (see Fig. 1(c)); these layers are AB stacked (see Fig. 1(d)) \cite{Ni3In_ED} within a unit cell (see Fig. 1(e)).  Density functional theory (DFT) calculations (see Fig. 1(f) with Brillouin zone in Fig. 1(g)) indicate that this structure supports a flat band within the $\mathrm{\Gamma}-\mathrm{M}-\mathrm{K}-\mathrm{\Gamma}$ ($k_z=0$) plane; among its large family of isostructural compounds, the flat band of \ce{Ni3In} is expected to appear at the Fermi level \cite{ICSD,MaterialsProject}. In Fig. 1(h) we illustrate the in-plane landscape of the flat band in \ce{Ni3In} (the bandwidth is approximately 60 meV), which itself can be traced downward in energy with $k_z$ (see Supplementary Materials).  Given that the Coulomb repulsion strength $U$ on the eV scale of elemental nickel \cite{Fe_Ni_DMFT} significantly outweighs the flat band width, we expect this to be a likely platform for correlated flat band states. 

We first examine the properties of \ce{Ni3In} over a broad range of temperature $T$, starting with the electrical resistivity in the kagome planes $\rho_{ab}(T)$.  As shown in Fig. 2(a), upon cooling from $T=300$ K, this is characterized by a broad shoulder which below $T \sim 100$ K gives way to approximately $T$-linear behavior.  A qualitatively similar response is observed for measurements out of the kagome plane ($\rho_{c}(T)$), with an overall reduction in magnitude and elevation in the resistivity exponent $\alpha$ in $\rho\sim T^{\alpha}$ ($\alpha$ for both traces is shown in the inset of Fig. 2(a) and is inferred from $\alpha=\partial\ln(\rho(T)-\rho_0)/\partial \ln T$ where $\rho_0$ is a constant representing the zero temperature limit of $\rho$).  This is a significant deviation from the conventional Fermi liquid behavior ($\alpha=2$) \cite{NFL_review} and bears a stark contrast to several structurally related kagome metals in which a flat band is not expected at $E_{F}$ (see Supplementary Materials).  Over the same range of $T$, the magnetic susceptibility $\chi(T)$ exhibits a prominent, non-saturating paramagnetism, with the high temperature $\chi_c$ being well-fit to a Curie-Weiss behavior (see Fig. 2(b); the reponse for $\chi_{ab}$ is similar, see Supplementary Materials).  The associated Weiss temperatures (effective magnetic moments per Ni) are $\theta_{c}=-50$ K (1 $\mu_B$) from $\chi_{c}$ and  $\theta_{ab}=-74$ K (1.2 $\mu_B$) from $\chi_{ab}$.  The lack of saturation in $\chi(T)$ together with the resistivity are suggestive of a non-Fermi liquid (NFL) phase at low $T$. At high $T$, phenomenologically $\rho_{ab}$ scales with $\chi_c$ down to approximately 80 K (see the inset of Fig. 2(b)), implying a magnetic origin of the underlying scattering process in the system, as we return to below.

The magnetoresistance (MR) response $\Delta \rho_{ab} \equiv \rho_{ab}(H) - \rho_{ab}(H=0)$ is shown in Fig. 2(c). The dominant response is a negative MR that grows rapidly with decreasing $T$, which we ascribe to field-induced suppression of magnetic fluctuations akin to that observed in magnetic metals above their ordering temperatures \cite{sd_scattering}. At the lowest $T$ and highest $H$ a positive curvature appears which likely reflects Lorentz-force induced MR.  For $T$ above which this orbital component is quenched, we find the MR can be scaled against $\mu_0H/(T+T^*)$ with $T^*=8.5$ K (see Fig. 2(d)).  This form of scaling was first applied to the experimental data of heavy fermion compound \ce{UBe13} with $T^*$ comparable with the Kondo temperature characterizing the antiferromagnetically coupled local moments and conduction electrons \cite{UBe13}, suggesting a local moment component of the fluctuating magnetism in \ce{Ni3In}.

Focusing on the low $T$ behavior, we show the specific heat normalized by temperature $C_{p}T^{-1}$ versus $T^2$ in Fig. 3(a).  Below approximately 15 K an upturn is observed, deviating from the form $\gamma+\beta T^2$ expected for a Fermi liquid ($\gamma$ is the Sommerfeld coefficient and $\beta T^2$ the phonon contribution).  Also shown is the conventional Fermi liquid response of the isostructural system \ce{Ni3Sn} which according to DFT exhibits a similar electronic structure apart from an overall energy shift of approximately 0.25 eV (moving the flat band away from $E_{F}$, see Supplementary Materials).  Using the acoustic phonon contribution of \ce{Ni3Sn}, for \ce{Ni3In} we find $\gamma \approx$ 47 mJ$\cdot$K$^2\cdot$mol$^{-1}$, an approximately a ten-fold increase from 5 mJ$\cdot$K$^2\cdot$mol$^{-1}$ for \ce{Ni3Sn}, consistent with that expected energetic shift of the flat band towards $E_F$ in \ce{Ni3In}.  From $\gamma$ we infer a density of states $\mathcal{D}(E_{F})=$ 40 eV$^{-1}$ u.c.$^{-1}$ for \ce{Ni3In}, nearly three times that estimated from DFT (14 eV$^{-1}$ u.c.$^{-1}$), signifying a considerable renormalization of the electronic states at $E_{F}$. 

High resolution measurements of $\rho_{ab}(T, H)$ offer an additional view into the low $T$ NFL behavior of \ce{Ni3In} (see Fig. 3(c)).  Seen most clearly in a map of $\alpha(T, H)$ in Fig. 3(d), the dark blue region ($\alpha\sim0.9$) near $H=0$ highlights a sublinear behavior of $\rho_{ab}$. Increasing $H$ tends to suppress the sub-linearity and drive the system towards a Fermi liquid state at low $T$. Such a sublinearity \cite{CeCoIn5_power} and $H$-driven suppression of NFL behavior (the latter also observed as maximum in $-d\chi/dT$ marked in Fig. 3(d) \cite{YFe2Al10}, see Supplementary Materials), recalls tuning of the $f$-electron non-Fermi liquid superconductor \ce{CeCoIn5} away from a quantum critical point above a characteristic field $H^*$  \cite{CeCoIn5_JP,CeCoIn5_WF}. There, the sublinear behavior was ascribed to the characteristic temperature scale of an underlying antiferromagnetic interaction \cite{CeCoIn5_JP,CeCoIn5_WF}; we hypothesize that a similar scenario may be at play here in \ce{Ni3In} with $H^*$ near zero.  At lower $T \lesssim 1$ K we observe an additional crossover in $\rho_{ab}(T)$ to $\alpha=2$ but no sign of a phase transition to an ordered state (see Supplementary Materials).  This metallic state is highly unusual, marked by an extreme anisotropy in the Kadowaki-Woods ratio $A/\gamma^2$ ($A$ is the coefficient of the $\alpha=2$ response): $A_{ab}/\gamma^2 = 226$ $\mu \Omega$ cm mol$^2$ K$^2$ J$^{-1}$ is among the largest ever reported, whereas $A_{c}/\gamma^2=0.65$ $\mu \Omega$ cm mol$^2$ K$^2$ J$^{-1}$ agrees well with that in elemental transition metals \cite{KW,KW2}, indicating an unconventional scattering processes taking place primarily within the kagome lattice plane. 

Based on these observations, we propose the behavior of \ce{Ni3In} can be understood in terms of an interplay between localized magnetic moments in the system with the conduction electron sea, where the former act as a primary source of scattering for the latter. A natural source of such magnetic moments are the flat band states, an identification corroborated by evidence of high temperature preformed local moments from the flat bands in local susceptibility calculations (see Supplementary Materials). Illustrated schematically in Fig. 3(b), at high $T$ the magnetic moments in the system are weakly coupled and exhibit a Curie-Weiss behavior as well as an onset of $H$-suppression of fluctuations.  Below approximately 80 K, short range correlations between the moments develop, leading to a growth of $\chi(T)$ deviating from a mean field Curie-Weiss form, and intense, anisotropic scattering of the conduction electrons including an approximate $T$-linear behavior for $\rho_{ab}(T)$. At lower $T \lesssim 20 $ K, the fluctuations become the dominant energy scale leading to NFL behavior apparent in all measured quantities.  Finally, for $T < 1$ K an enigmatic metallic state appears marked by an extreme anisotropy of scattering enhanced in the kagome plane. To shed light on any proximate magnetic order that can serve as a source of quantum fluctuations generating the observed scattering processes in \ce{Ni3In}, we performed theoretical analysis of the momentum($\bm{q}$)-dependent $\chi(\bm{q})$ and found an enhanced response within the flat band plane which maximizes in regions enclosing the $\mathrm{M}$ point (see Supplementary Materials).

We further explore the flat band states in \ce{Ni3In} by means of angle-resolved photoemission spectroscopy (ARPES). While difficult to capture at $E_{F}$ (see Fig. 4(a,b) the $k_{z}=0$ plane $\mathrm{\Gamma}-\mathrm{K}-\mathrm{M}-\mathrm{\Gamma}$), the flat band is expected to disperse along $\mathrm{H}-\mathrm{K}-\mathrm{H}$ as is observed in Fig. 4(c,d).  By renormalizing the DFT bands by a factor of 0.7, we find reasonable agreement with the overall dispersive features from ARPES, suggesting that the flat band placed at $E_F$ by DFT sets an adequate starting point for the understanding of the unusual transport and thermodynamic responses observed in \ce{Ni3In}.  In Fig. 4(e) we show the orbitally projected $\mathcal{D}(E)$, which identifies the $d_{xz}$ orbital as the origin for the flat band and the primary source of $\mathcal{D}(E)$ at $E_F$ (note the orbital axes are defined locally, see Fig. 4(e) inset).  The fundamental unit of the flat band wave function is found to be a Wannier state localized on the Ni triangular plaquette (see Fig. 4(f-h)). This naturally explains the appearance of a flat band as a result of a destructive interference between nearest (intra-plane, blue arrow in Fig. 4(h)) and next-nearest (inter-plane, red arrow in Fig. 4(h)) hopping processes between the plaquettes (see the tight-binding modeling on the bilayer kagome lattice in Supplementary Materials), which arises from the alternating sign pattern of the $d$-orbital texture. Along the $z$-direction, a considerable overlap between the $d_{xz}$ orbital lobes (see Fig. 4(f)) is consistent with the strong $k_z$ dispersion of the  band observed in ARPES (Fig. 4(c)), the highly anisotropic transport response, and the peak in $\mathcal{D}(E_{F})$ (expected for the $k_{z}=0$ plane of a 1D dispersion).   

To understand the consequence of electronic correlations, it is instructive to compare the $d_{xz}$ flat band to that in the original $s$-orbital model on the kagome lattice \cite{Balents_FB} in which destructive interference (illustrated as blue and red arrows in Fig. 4(i)) arises from the anti-bonding nature of the wave function trapped in hexagonal plaquettes. These support distinct networks for the local wave functions: the overlapping nature of the flat band wavefunction on adjacent hexagons (Fig. 4(i)) for the $s$-orbital model has been suggested to drive a collective ferromagnetism \cite{Mielke_Tasaki} while the non-overlapping wave function illustrated in Fig. 4(h) resembles a lattice of isolated $f$-electron states with a tendency for local moment formation \cite{Si_review} (a distinction we refer to as delocalized and localized flat bands). Given that placing a lattice of local moments within a conduction electron sea has long been known as the fundamental construction of $f$-electron-based NFL systems \cite{Si_review}, we anticipate this scheme may also be the key to understanding NFL in \ce{Ni3In}, while the salient differences between systems with $f$-electrons and the present $d$-electron flat band including anisotropy and the spatial extent of wave function \cite{PeriodicTable} may enable new avenues of constructing quantum criticalities in metallic systems.

Tuning \ce{Ni3In} across a potential quantum phase transition should help reveal the ultimate fate of the flat band electrons in the system in the presence of correlation. From the materials perspective, \ce{Ni3In} belongs to a host crystalline structure that includes over 200 materials entries in ICSD \cite{ICSD}, within which the diversity of transition metal, rare earth, and $p$ block elements may enable highly flexible chemical tuning of correlation of the flat band states.  It is noteworthy that among these materials and related structures there are a number of topologically non-trivial kagome metals including non-collinear antiferromagnetic \ce{Mn3(Sn,Ge)} \cite{Mn3Sn,Mn3Ge}, ferromagnetic \ce{Fe3Sn2} \cite{Fe3Sn2}, collinear antiferromagnetic FeSn \cite{FeSn}, and superconducting and charge density wave supporting \ce{CsV3Sb5} \cite{CsV3Sb5},  suggesting multiple pathways to engineer both the symmetry-breaking ground state and topology of an ordered state proximate to the NFL behavior in \ce{Ni3In}. The nickel triangular plaquettes in \ce{Ni3In} itself form a triangular lattice, which in the context of antiferromagnetic interaction may lead to strong geometric frustration and emergent spin excitations relevant to understanding the present NFL behavior \cite{triangularlattice,CePdAl}. Our results provide a proof-of-principles example that using a crystalline material from the lattice and orbital engineering point of view, a flat band may be brought to $E_{F}$ to create a correlated metallic state. We anticipate this approach can also be applied to enhance electronic correlation in materials based on elements not limited to the paradigmatic $d$ and $f$ blocks \cite{PeriodicTable} and serve as a stepping stone to future discoveries of novel topologically non-trivial many-body states that derive from massive flat band degeneracies \cite{Sun_FB,Wen_FB,Mudary_FB}.

\textit{Note} After completion of this work, we became aware of a recent database search of flat band materials \cite{FB_catalogue}. It is of significant interest to combine the present lattice/orbital engineering approach with such high-throughput catalogs to efficiently investigate new crystalline flat band materials.

\vspace{10 mm}

\textbf{Acknowledgments} We appreciate fruitful discussions with T. Senthil, B.J. Yang, L. Zou, Y. Zhang, K. Haule, J.-S. You and J. van den Brink. This research is funded in part by the Gordon and Betty Moore Foundation through Grants GBMF3848 and GBMF9070 to J.G.C., NSF grant DMR-1554891, and ARO Grant No. W911NF-16-1-0034. L.Y., M.K. and E.K. acknowledge support by the STC Center for Integrated Quantum Materials, NSF grant number DMR-1231319.  L.Y. acknowledges the Heising-Simons Physics Research Fellow Program and the Tsinghua Education Foundation. S.F. is supported by a Rutgers Center for Material Theory Distinguished Postdoctoral Fellowship. M.K. acknowledges support from the Samsung Scholarship from the Samsung Foundation of Culture. R.C. acknowledges support from the Alfred P. Sloan Foundation. O.J. was supported by the Leibniz Association through the Leibniz Competition. A portion of this work was performed at the National High Magnetic Field Laboratory, which is supported by the National Science Foundation Cooperative Agreement no. DMR-1644779, the State of Florida and the DOE.  This research used the resources of the Advanced Light Source, a US Department of Energy (DOE) Office of Science User Facility under contract no. DE-AC02-05CH11231. The computations in this paper were run on the ITF/IFW computer clusters (Dresden, Germany) the FASRC Cannon cluster supported by the FAS Division of Science Research Computing Group at Harvard University. We thank U. Nitzsche for technical assistance in maintaining computing resources at IFW Dresden.


\section{Methods}

\textbf{Materials Synthesis} Single crystals of \ce{Ni3In} are synthesized via a catalyzed reaction. The polycrystals of Ni$_3$In$_{1-x}$Sn$_x$ are synthesized via a solid state reaction. The phase of obtained crystals is confirmed with X-ray diffraction.

\textbf{Physical Properties Measurements} The electrical transport measurements were performed on single crystals and polycrystals with the standard five probe method in a commercial cryostat and also at the National High Magnetic Field Laboratory DC field facility. The heat capacity measurements are performed on sintered polycrystals using the two relaxation time method. Magnetic susceptibility measurements are performed in the Vibrating Sample Magnetometer (VSM) with the Quantum Design Magnetic Property Measurement System (MPMS3). 

\textbf{Scanning Transmission Electron Microscopy} Scanning Transmission Electron Microscopy (STEM) experiments were conducted on a JEOL ARM 200CFG probe corrected microscope at an accelerating voltage of 200 kV.  \ce{Ni3In} samples were prepared by a Helios focused-ion beam (Thermo Electron) operated at an acceleration voltage of 30 kV for the gallium beam lift-out, followed by 1 keV final Argon polish with a (Fischione) Nanomill for 15 minutes.

\textbf{Angle-resolved Photoemission Spectroscopy} Angle-resolved Photoemission Spectroscopy (ARPES) experiments were performed at the Beamline 7.0.2 (MAESTRO) of the Advanced Light Source. The experiments were conducted at the micro-ARPES endstation equipped with a R4000 hemispherical electron analyzer (Scienta Omicron). ARPES measurements were conducted at the liquid nitrogen temperature ($\sim80$ K) and under the ultrahigh vacuum better than $4\times10^{-11}$ torr. Photon energy-dependence was investigated over a wide energy range from 70 eV to 230 eV to identify high-symmetry positions along the $k_z$ momentum-space directions. The high-symmetry points ($\Gamma$ and Z) and periodicity along $k_z$ were well-reproduced through the nearly-free-electron final state model with inner potential 10 eV. All spectra were collected with $p$-polarized photons. The energy and momentum resolutions were below 20 meV and 0.01 $\text{\AA}^{-1}$, respectively.

\textbf{First-principles Electronic Structure Calculation} The \textit{ab initio} density functional theory (DFT) calculations were performed by Vienna Ab initio Simulation Package (VASP) \cite{vasp1,vasp2}. The pseudo potential formalism is based on the Projector augmented wave method (PAW) \cite{PAW} with exchange-correlation energy parametrized by Perdew, Burke and Ernzerhof (PBE)~\cite{pbe}, a functional of the generalized gradient approximation (GGA) type. The DFT calculations for the bulk \ce{Ni3In} crystal are converged with an energy cut-off 360 eV for the plane-wave basis and a $13\times13\times11$ Monkhorst-Pack grid sampling in the reciprocal space. To analyze the DFT electronic structure, we employed the Wannier90 code~\cite{mlwf,mlwf_new} to construct Wannier tight-binding Hamiltonian~\cite{wannier_review}, using the localized Wannier basis projected from Ni $3d$, $4s$ and In $5s$ states. The further simplified effective 6-band model projected from the locally rotated $d_{xz}$ orbitals is also derived similarly.

To construct an effective model based on molecular rather than atomic orbitals, we additionally carried out DFT calculations using the full-potential local-orbital code FPLO~\cite{fplo} version 18. Its built-in module~\cite{fplo-wf} allows the user to construct Wannier projections for molecular-like states comprising several atomic orbitals. Since the GGA orbital-resolved density of states (DOS) revealed the dominance of Ni $d_{xy}$ and $d_{xz}$ contributions in the vicinity of the Fermi level, we restricted ourselves to these atomic orbitals. Molecular orbitals were constructed by combining the respective orbitals of three Ni atoms forming a triangle on the kagome lattice. In this way, we obtained a effective model with two sites per cell and two orbitals per site, giving rise to four bands. To cross-check results magnetic susceptibilities calculated using different models, we also performed an automatic wannierization, which yields a model describing all valence states (excluding core and semi-core states).

\pagebreak
\begin{figure}[h]
	\includegraphics[width = 0.7  \columnwidth]{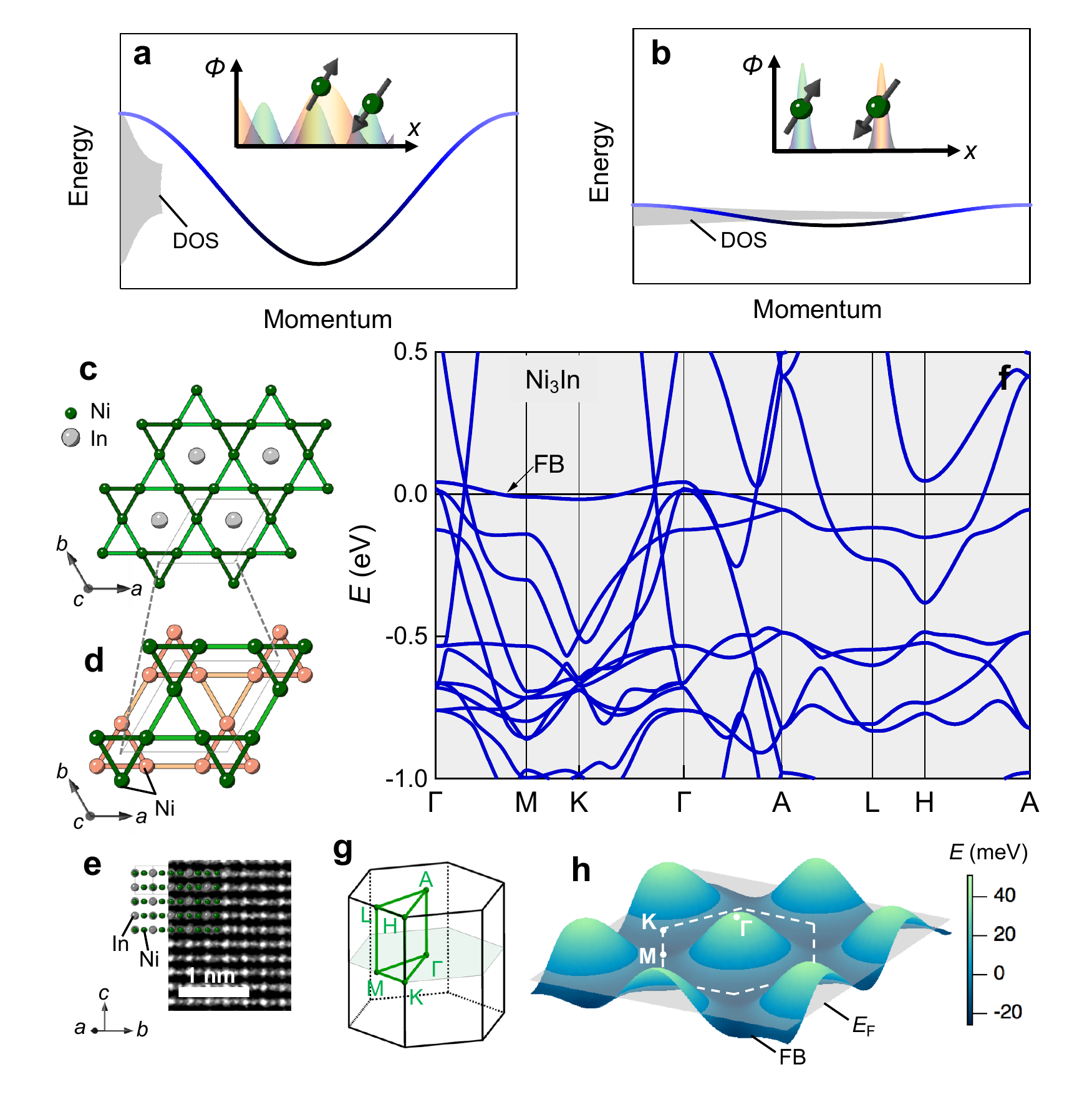}
	\caption{\label{fig-1} \textbf{The kagome lattice structure and flat band in \ce{Ni3In}} (a) Schematic of the momentum dispersion of a prototypical Bloch band.  The shaded area represents the associated density of states (DOS). The inset illustrates the corresponding Bloch wave functions in real space. (b) Schematic of a flat band with a concentrated DOS near the flat band energy. The inset shows spatially localized flat band wave functions. (c) Top view of an isolated \ce{Ni3In} layer in the $ab$ plane.  Green and gray atoms represent nickel and indium, respectively. The dark green bonds (2.510  $\text{\AA}$) are shorter than light green bonds (2.842  $\text{\AA}$), which together define a breathing kagome lattice. (d) Top view of the nickel sublattice in \ce{Ni3In}.  Each unit cell is composed of A-B stacked kagome bilayers of Ni. Ni atoms on the top and bottom layers are colored as green and pink, respectively. (e) Transmission electron microscopy (TEM) image viewed from [210] direction overlaid with the corresponding cross section of the crystal structure. The scale bar stands for 1 nm. (f) Density functional theory (DFT) band structure of \ce{Ni3In} without spin-orbit coupling. The high symmetry points and lines in the Brillouin zone are highlighted in (g). (h) Flat band (FB) dispersion with in $k_z=0$ plane illustrated along with the Fermi level $E_F$.}
\end{figure}

\pagebreak
\begin{figure}[h]
	\includegraphics[width = 0.8 \columnwidth]{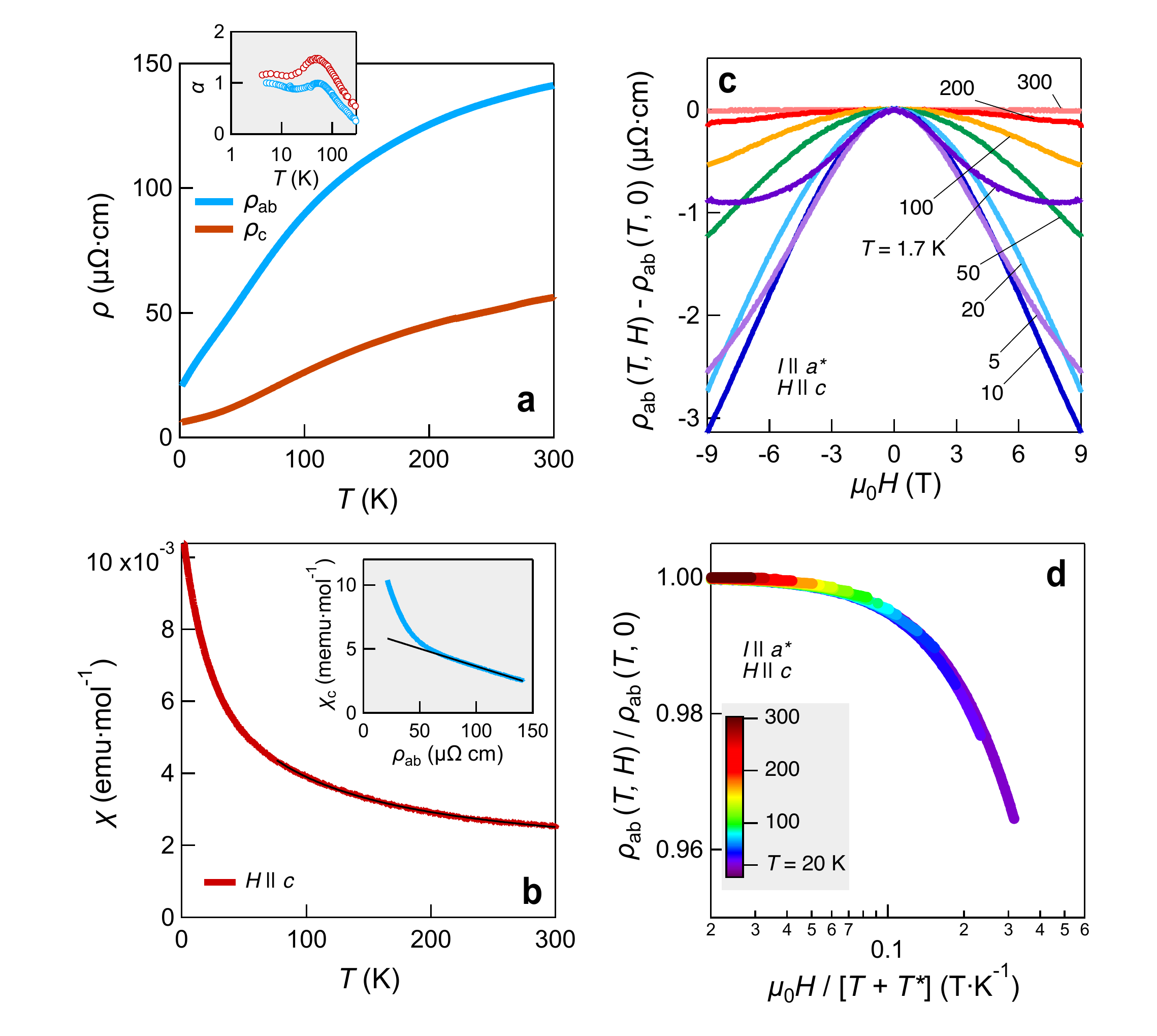}
	\caption{\label{fig-2} \textbf{Non-Fermi liquid behavior and magnetotransport in \ce{Ni3In}} (a) Resistivity $\rho$ of \ce{Ni3In} as a function of temperature $T$ in (blue) and out of (red) the kagome lattice plane. The inset shows resistivity exponent $\alpha$ (see text) for both $\rho_{ab}$ (blue symbols) and $\rho_c$ (red symbols). (b) Magnetic susceptibility $\chi$ of Ni$_3$In single crystal with magnetic field out of plane. At high $T$, $\chi$ may be described by a Curie-Weiss fit (solid line). The inset plots $\chi_c$ against $\rho_{ab}$ and the solid line represents a linear fit. (c) $\Delta\rho_{ab}=\rho_{ab}(T,H)-\rho_{ab}(T,0)$ as a function of magnetic field $\mu_0H$ at selected $T$.  (d) Negative magnetoresistance $\rho_{ab}(T,H)/\rho_{ab}(T,0)$ shown with magnetic field normalized by $T+T^{\star}$ (see text).}
\end{figure}

\pagebreak
\begin{figure}[h]
	\includegraphics[width = 0.8 \columnwidth]{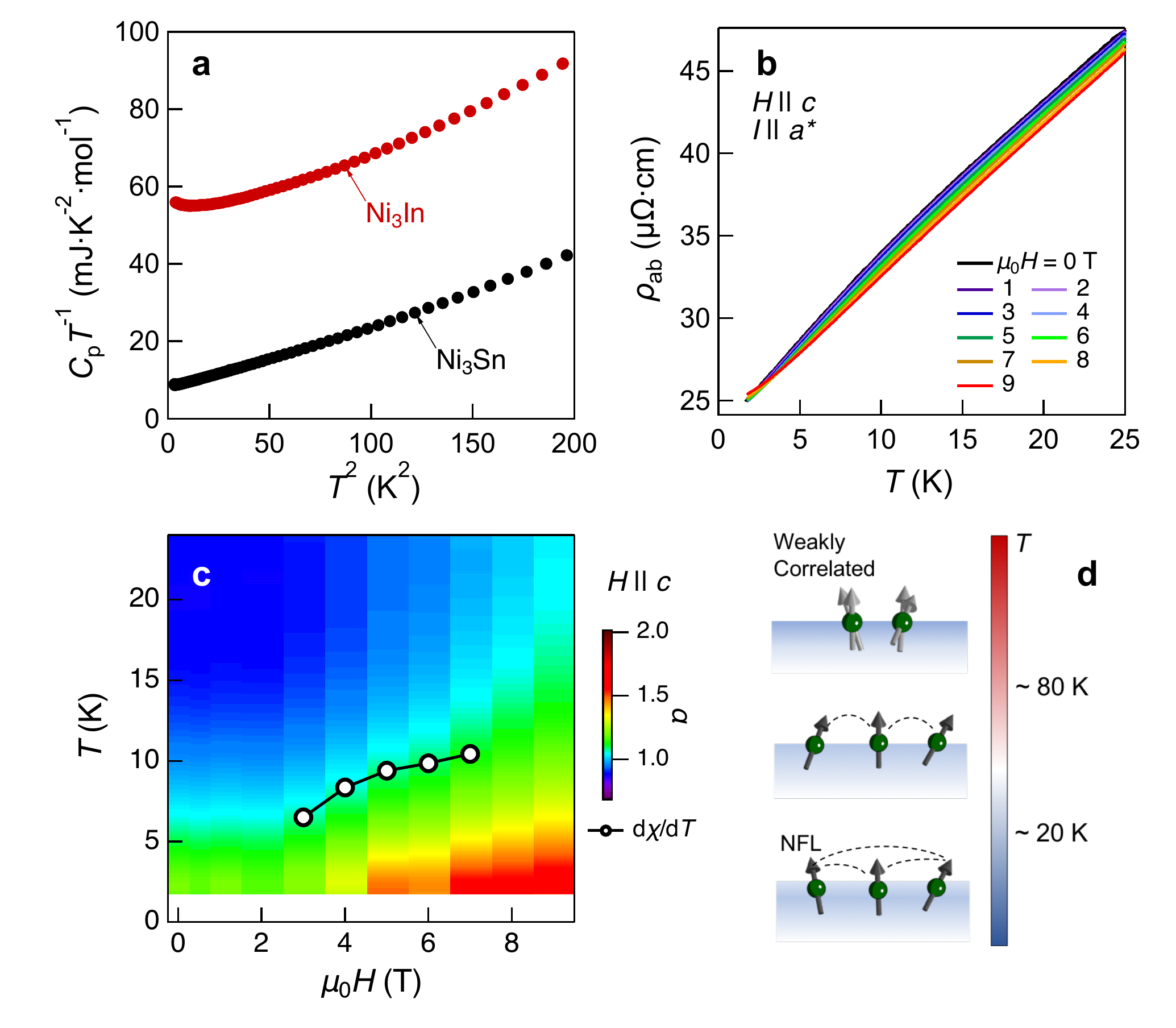}
	\caption{\label{fig-3} \textbf{Low $T$ non-Fermi liquid behavior in \ce{Ni3In}} (a) Heat capacity normalized by temperature $C_p\cdot T^{-1}$ of \ce{Ni3In} (red symbols) and isostructural \ce{Ni3Sn} (black symbols) with respect to $T^2$. (b) In-plane resistivity below 25 K at various applied magnetic fields along the $c$-axis. (c) Resistivity power law $\alpha$ in the temperature/magnetic field phase space. The circles are taken as maximum of $d\chi/dT$ (see Supplementary Materials). (d) Schematic of the crossover of the system from a high temperature regime with weakly correlated localized magnetic moments to a low temperature non-Fermi liquid (NFL) regime with fluctuations of potentially nearby magnetic orders. The blue shade represents the conduction electron sea.
	}
\end{figure}

\pagebreak
\begin{figure}[h] 
	\includegraphics[width = 0.65 \columnwidth]{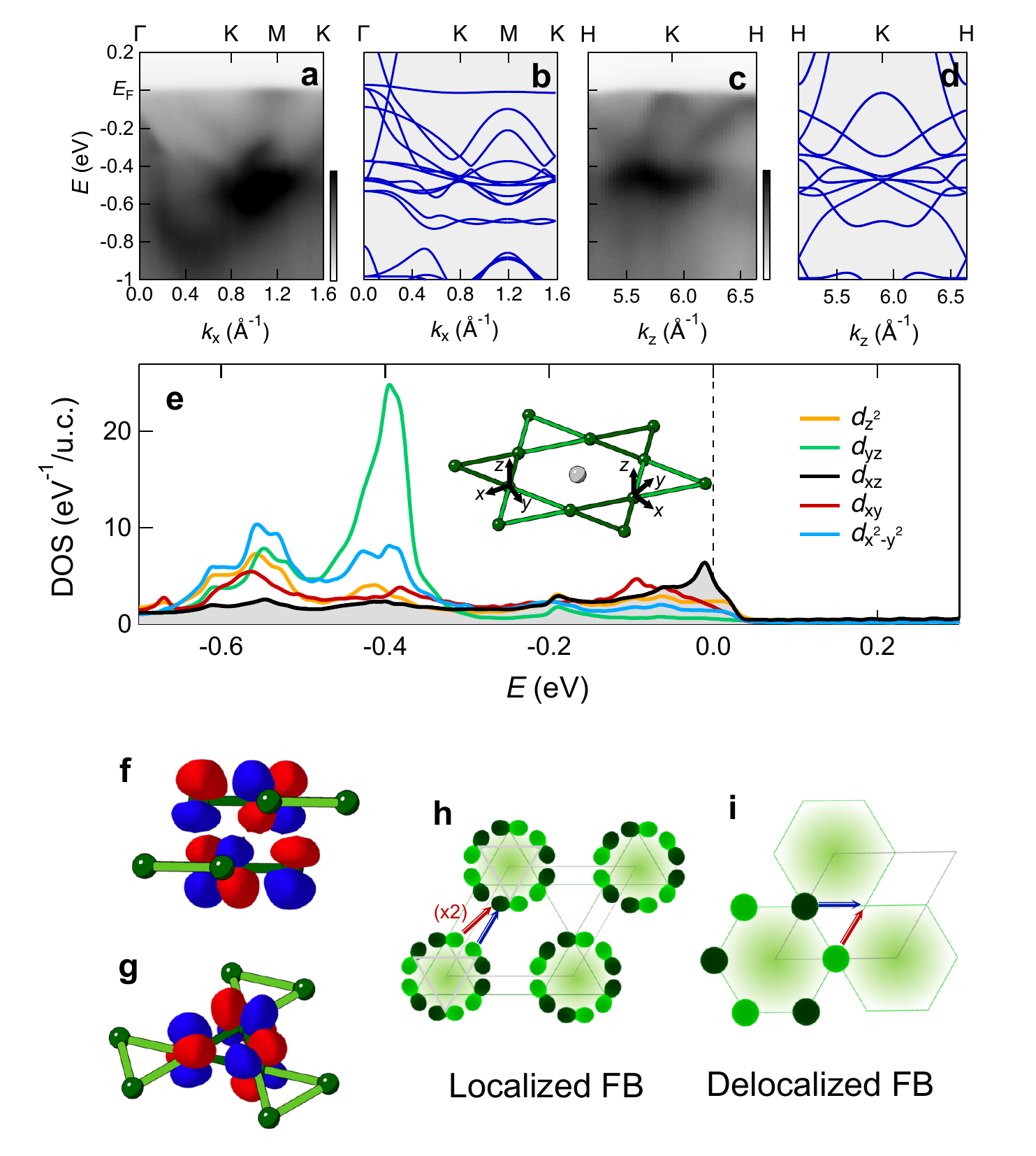}
	\caption{\label{fig-4} \textbf{The flat band wave function in \ce{Ni3In}} (a-d) Angle-resolved photoemission spectroscopy (ARPES) spectra measured at 20 K of \ce{Ni3In} (a,c) contrasted with corresponding DFT band structure (b,d). (a,b) are along an in-plane $\mathrm{\Gamma}-\mathrm{K}-\mathrm{M}-\mathrm{\Gamma}$ momentum cut and (c,d) are along an out-of-plane $\mathrm{H}-\mathrm{K}-\mathrm{H}$ momentum cut. The DFT band dispersions in (b) and (d) are renormalized by a factor of 0.7. (e) Density of states spectrum decomposed into respective $d$ orbital degrees of freedom. As schematically illustrated in inset, for each nickel site the local $x$-axis points to the In atom in the kagome lattice plane and $y$-axis points to the center of each nickel triangle. The $z$-axis aligns with the crystallographic $c$-axis. (f) Side and (g) three-dimensional view of the Wannier function of the $k_z=0$ flat band in \ce{Ni3In}, which is composed of local $d_{xz}$ orbitals. (h) Top view of multiple localized flat band states of \ce{Ni3In} on the lattice. (i) Schematic of the flat band wave function for an $s$-orbital kagome lattice. In (h) and (i) dark and light green indicate positive and negative portions of the flat band wave functions, respectively, and red and blue arrows represents hopping paths with opposite sign. The shades in (h,i) highlight the spatial distribution of the flat band wave function within a fundamental unit in each case.}
\end{figure}

\pagebreak

\end{document}